\documentclass[11pt,a4paper]{article}
\usepackage{epsfig}
\RequirePackage{mathrsfs}

\newlength{\dinwidth}
\newlength{\dinmargin}
\def\eq#1{{Eq.~(\ref{#1})}}
\newcommand{\Le}{\left(}
\newcommand{\Ra}{\right)}

\newcommand{\beq}{\begin{equation}}
\newcommand{\eeq}{\end{equation}}
\newcommand{\beqar}{\begin{eqnarray}}
\newcommand{\eeqar}{\end{eqnarray}}
\begin{document}

\title {{~}\\
{\Large \bf  Mean field approximation for the dense charged drop }\\}
\author{ 
{~}\\
{~}\\
{\large 
S.~Bondarenko$\,\,$\thanks{\,The author declares that there is no conflict of interest regarding the publication of this paper.},
K.~Komoshvili$\,\,$\thanks{\,The author declares that there is no conflict of interest regarding the publication of this paper.}} 
\\[10mm]
{\it\normalsize  Ariel University, Israel}\\}

\maketitle
\thispagestyle{empty}

\begin{abstract}
In this note, we consider the mean field approximation for the description of the probe charged particle in a dense charged drop.
We solve the corresponding Schr\"{o}dinger equation for the drop with spherical symmetry in the first order of mean field  approximation
and discuss the obtained results.

\end{abstract}

\newpage
\section{Introduction}

\,\,\,\,Collisions of relativistic nuclei in the RHIC and LHC experiments at very high energies led to the discovery of a new state of matter named quark-gluon plasma (QGP). At the 
initial stages of the scattering, this plasma resembles almost an ideal liquid 
 whose microscopic structure is not yet well understood \cite{Shyr1,Shyr2,Berd,Koch,Liao,Nahrgang,Stein,Skokov,Rand}. The data obtained in the RHIC experiments is in a good agreement with the predictions of the 
ideal relativistic fluid dynamics, \cite{Fluid,Fluid1,Fluid2}, that establishes fluid dynamics as the main theoretical tool
to describe collective flow in the collisions. As an input to the hydrodynamical evolution of the particles it
is assumed that after a very short time, $\tau\, <\, 1\,\, fm/s$ \cite{Fluid2},  the matter  reaches a thermal equilibrium and expands with a very small shear viscosity  \cite{Visc,Teaney}.

   In this paper we continue to develop the model proposed in \cite{Our}. Namely, we assume that at the local energy density fluctuations,  hot drops \cite{HotSp,HotSp1,Step}, are created at the very initial stage of interactions at  times $\tau\,<\,0.5\,fm\,$. These fireballs (hot drops) are very dense and small, their size is much smaller than the proton size, 
see  \cite{Our}, and their energy density is much larger than the density achieved  in high-energy interactions at the energies $\sqrt{s}\,<\,100\,GeV\,$. In our model, we assume that the fireballs  
consist of the particles with weak inter-particle interactions and have a non-zero charge.
This drop of charged particles we consider from the point of view of quantum statistical physics.
The most general Hamiltonian for this system can be written in the form which describes all possible interactions between the particles in the drop:
\beq\label{1Ch01}
H\,= \,H^{0}\,+\,V^{1}\,+\,V^{2}\,+\,\ldots\,+V^{i}\,+\,\ldots\,,
\eeq
see \cite{Land}, where as usual $V^{1}$ is an energy of interaction of the particle with an external field, the $V^{2}$ is the energy of pair-like interactions and etc.
The mean field approximation for the probe particle in the system of charged particles, therefore, can be introduced by the following perturbative scheme. 
First of all, we can consider the motion of only one probe particle in the mean field of all other particles, that corresponds to preserving only $V^{1}$ term in the \eq{1Ch01} expression.
This approximation will lead to the modification of the propagator of the particle, namely from a free propagator to some "dressed" one.
At the next step, we can take two probe particles, each of them will propagate in the mean field of the other charges of the system, similarly to the first approximation, 
but additionally we can introduce the interaction of these two particles one with another in the mean field of the remaining charges in the drop, 
that requires introduction of one $V^{2}$ term in \eq{1Ch01} expression in the mean field approximation. 
Further, we can increase the number of the probe particles in the system, considering, additionally to pair interactions, the interactions of free probe particles  and so on.

  In present calculations, we limit ourselves to the first order of the mean field approach, 
namely we will consider the motion of one non-relativistic probe particle
in the external mean field created by all other particles in the charged drop.

\section{Mean field approximation for the Hamiltonian of the system }

 In the absence of the external field, we write the Hamiltonian of the system of charged particles as
\beqar\label{1Ch1}
H\, & = & \,-\,\frac{1}{2\,m}\,\int\,\Psi_{\alpha}^{+}(t,\,\vec{r})\,\Delta\,\Psi_{\alpha} (t,\,\vec{r})\,d^{3} x \,-\mu\,N+\,\nonumber \\
& + & \, \frac{1}{2}\,\int\,\Psi_{\beta}^{+}(t,\,\vec{r})\,\Psi_{\alpha}^{+} (t,\,\vec{r}^{\,'})\,U(\vec{r}\,-\,\vec{r}^{\,'})\,
\Psi_{\alpha}(t,\,\vec{r}^{\,'})\,\Psi_{\beta} (t,\,\vec{r})\,d^{3} x\,d^{3} x^{'}\,,
\eeqar
here
\beq\label{1Ch2}
N\,=\,\int\, \Psi_{\alpha}^{+}(t,\,\vec{r})\,\Psi_{\alpha} (t,\,\vec{r})\,d^{3} x 
\eeq
is a particles number operator.
Considering the mean field approximation for spherically symmetrical system, we introduce
\beq\label{1Ch3}
\Psi_{\alpha}^{+} (t,\,\vec{r}^{\,'})\,\Psi_{\beta} (t,\,\vec{r})\,\approx\,<\,\Psi_{\alpha}^{+} (t,\,\vec{r}^{\,'})\,\Psi_{\beta} (t,\,\vec{r})\,>\,=\,\delta_{\alpha\,\beta}\,
f(r,\,r^{'},\,r_{0})
\eeq
as some particles density for the droplet with characteristic size $r_{0}$, here $r\,=\,|\vec{r}|$. The $f(r,\,r^{'},\,r_{0})$ function is a distribution function of the system of interests, it can be correctly determined by 
writing corresponding Vlasov's or Boltzmann's  equations coupled to \eq{1Ch01} system. In our case, we will not consider a particular form of this function, instead we will discuss it's form basing on some 
physical assumptions only. Therefore,
we obtain for the Hamiltonian:
\beqar\label{1Ch4}
H\, & = & \,-\,\frac{1}{2\,m}\,\int\,\Psi_{\alpha}^{+}(t,\,\vec{r})\,\Delta\,\Psi_{\alpha} (t,\,\vec{r})\,d^{3} x \,-\mu\,N+\,\nonumber \\
& + & \, \, \frac{1}{2}\,\int\,\Psi_{\alpha}^{+}(t,\,\vec{r})\,U(\vec{r}\,-\,\vec{r}^{\,'})\,f(r,\,r^{'},\,r_{0})\,,
\Psi_{\alpha}(t,\,\vec{r}^{\,'})\,\,d^{3} x\,d^{3} x^{'}\,,
\eeqar
which represents now the energy of the probe particle in the mean field created by the other particles of the system.
Due to the  spherical symmetry of the problem, we expand all the operators in the Hamiltonian's expression in terms of spherical harmonic functions.
We have for the two-particles interaction potential:
\beq\label{1Ch5} 
U(\vec{r}\,-\,\vec{r}^{\,'})\,= \,\sum_{l\,=\,0}^{\infty}\,\sum_{m\,=\,-\,l}^{l}\,\frac{4\,\pi\,q^{2}}{2\,l\,+\,1}\,
\Le\theta(r^{\,'}\,-\,r)\,
\frac{r^{l}}{\Le r^{'}\Ra^{l+1}}\,+\,
\theta(r\,-\,r^{\,'})\,\frac{r^{'\,l}}{r^{l+1}}\,\Ra\,Y_{l m}^{*}(\Psi,\,\Phi)\,Y_{l m}(\psi,\,\phi)\,
\eeq
with $\Psi,\,\Phi$ as spherical angles of $r^{'}$ vector, $\psi,\,\phi$ as spherical angles of $r$ vector in some spherical coordinate system and
$\theta(r)$ as the step function.
Correspondingly,
we write the particle-field operator as
\beq\label{1Ch6}
 \Psi_{\alpha} (t,\,\vec{r})\,=\,\sum_{l\,=\,0}^{\infty}\,\sum_{m\,=\,-\,l}^{l}\,\psi_{\alpha\,l\,m}(t,\,r)\,Y_{l m}(\psi,\,\phi)\,.
\eeq
Using the orthogonality property of the harmonic functions
\beq\label{1Ch7}
\int\,Y_{l m}(\psi,\,\phi)\,Y_{l^{'} m^{'}}^{*}(\psi,\,\phi)\,d\, \Omega\,=\,\delta_{l\, l^{'}}\,\delta_{m\, m^{'}}\,,
\eeq
with $d^{3} x\,=\,r^{2}\,d r\,d\,\Omega$, we rewrite the Hamiltonian \eq{1Ch4} in one-dimensional form as a function of $r$ and $r^{'}$ only:
\beqar\label{1Ch8}
H& = &- \frac{1}{2\,m}\sum_{l\,=\,0}^{\infty}\sum_{m\,=\,-\,l}^{l}\int_{0}^{\infty}\Le\,\psi_{\alpha\,l\,m}^{+}(t,\,r)\,
\Delta_{r}\,\psi_{\alpha\,l\,m}(t,\,r)\,-\,\frac{l\,(l\,+\,1)}{r^{2}}\,\psi_{\alpha\,l\,m}^{+}(t,\,r)\,\psi_{\alpha\,l\,m}(t,\,r)\Ra\,r^{2}\,dr\,\,-\,\nonumber \\
\,& - &\,\mu\,\sum_{l\,=\,0}^{\infty}\sum_{m\,=\,-\,l}^{l}\int_{0}^{\infty}\,\psi_{\alpha\,l\,m}^{+}(t,\,r)\,\psi_{\alpha\,l\,m}(t,\,r)\,r^{2}\,d r\,+\nonumber\\
& + &\,\sum_{l\,=\,0}^{\infty}\,\sum_{m\,=\,-\,l}^{l}\,\frac{4\,\pi\,q^{2}}{2\,l\,+\,1}\,
\int_{0}^{\infty}\,r^{2}\,d r\,\Le\,\int_{r}^{\infty}\,\frac{r^{l}}{r^{'\,(\,l\,+\,1\,)}}\,f(r,\,r^{'},\,r_{0})\,
\psi_{\alpha\,l\,m}^{+}(t,\,r)\,\psi_{\alpha\,l\,m}(t,\,r^{'})\,r^{'\,2}\,d r^{'}\,+\,\right.\nonumber \\
&+&\,\left.\,\int_{0}^{r}\,\frac{r^{'\,l}}{r^{(l\,+\,1)}}\,f(r,\,r^{'},\,r_{0})\,
\psi_{\alpha\,l\,m}^{+}(t,\,r)\,\psi_{\alpha\,l\,m}(t,\,r^{'})\,r^{'\,2}\,d\, r^{'}\,\Ra\,.
\eeqar
In the next Section, we solve Schr\"{o}dinger equation corresponding to this Hamiltonian.

\section{Equations of motion}

 We introduce the usual commutation relations for the fields of interest (see \eq{1Ch6}):
\beq\label{2Ch1}
\{\,\Psi_{\alpha} (t,\,\vec{r}),\,\Psi_{\beta}^{+} (t,\,\vec{r}^{\,'})\,\}\,=\,\delta_{\alpha\,\beta}\,\delta^{3}(\,\vec{t}\,-\,\vec{r}^{\,'})\,.
\eeq
Using \eq{1Ch7} property, we correspondingly obtain one-dimensional commutation relations for the new fields:
\beq\label{2Ch2}
\{\,\psi_{\alpha\,l\,m}(t,\,r),\,\psi_{\beta\,l^{\,'}\,m^{\,'}}^{+}(t,\,r^{\,'})\,\}\,=\,\frac{1}{r^{2}}\,\delta_{\alpha\,\beta}\,\delta_{l\, l^{'}}\,\delta_{m\, m^{'}}\,\delta(\,r\,-\,r^{\,'})\,.
\eeq
The Schr\"{o}dinger equation for $\psi_{\alpha\,l\,m}(t,\,r)$ field, therefore, has the following form:
\beqar\label{2Ch3}
\imath\,\frac{\partial }{\partial t}\,\psi_{\alpha\,l\,m}(t,\,r)\,& = &\,
\Le\,-\,\frac{1}{2 m}\,\Le\, \Delta_{r}\,-\,\frac{l\,(l\,+\,1)}{r^{2}}\,\Ra\,-\,\mu\,\Ra\,\psi_{\alpha\,l\,m}(t,\,r)\,+\,\nonumber \\
& + &\,
\frac{4\,\pi\,q^{2}}{2\,l\,+\,1}\,\Le
\int_{r}^{\infty}\,\frac{r^{l}}{r^{'\,(\,l\,+\,1\,)}}\,f(r,\,r^{'},\,r_{0})\,
\psi_{\alpha\,l\,m}(t,\,r^{'})\,r^{'\,2}\,d r^{'}\,+\,\right.\nonumber \\
&+&\,\left.\,\int_{0}^{r}\,\frac{r^{'\,l}}{r^{(l\,+\,1)}}\,f(r,\,r^{'},\,r_{0})\,
\psi_{\alpha\,l\,m}(t,\,r^{'})\,r^{'\,2}\,d\, r^{'}\,
\Ra\,.
\eeqar
Rescaling the drop's density function and rewriting it in the dimensionless form as 
\beq\label{2Ch4}
f(r,\,r^{'},\,r_{0})\,\rightarrow\,\frac{1}{r^{3/2}\,(r^{\,'})^{3/2}}\,f(r,\,r^{'},\,r_{0})\,,
\eeq
introducing new variable in integrals in \eq{2Ch3} 
\beq\label{2Ch5}
r^{\,'}\,=\,x\,r\,,
\eeq
we rewrite the integrals in \eq{2Ch3} finally as
\beq\label{2Ch6}
\frac{4\,\pi}{r\,\Le\,2\,l\,+\,1\,\Ra}\,\Le
\int_{1}^{\infty}\,\frac{d x}{x^{\,l\,+\,1/2}}\,f(r,\,x,\,r_{0})\,\psi_{\alpha\,l\,m}(t,\,x\,)\,+\,
\int_{0}^{1}\,d x\,x^{\,l\,+\,1/2}\,f(r,\,x,\,r_{0})\,\psi_{\alpha\,l\,m}(t,\,x\,)\,
\Ra\,.
\eeq
For the case of the drop of small size, we can expand our $\psi$ function in \eq{2Ch6} around $x\,=\,1\,$ in both terms, this point gives the main contribution to both integrals.
Therefore, in the first approximation, we have for \eq{2Ch6}:
\beqar\label{2Ch7} 
&\,&\frac{4\,\pi}{r\,\Le\,2\,l\,+\,1\,\Ra}\,\Le
\int_{1}^{\infty}\,\frac{d x}{x^{\,l\,+\,1/2}}\,f(r,\,x,\,r_{0})\,\psi_{\alpha\,l\,m}(t,\,x r\,)\,+\,
\int_{0}^{1}\,d x\,x^{\,l\,+\,1/2}\,f(r,\,x,\,r_{0})\,\psi_{\alpha\,l\,m}(t,\,x r\,)\,
\Ra\,\approx\,\nonumber \\
&\,& \psi_{\alpha\,l\,m}(t,\, r\,)\frac{4\,\pi}{r\,\Le\,2\,l\,+\,1\,\Ra}
\Le
\int_{1}^{\infty}\,\frac{d x}{x^{\,l\,+\,1/2}}\,f(r,\,x,\,r_{0})+
\int_{0}^{1}\,d x\,x^{\,l\,+\,1/2}\,f(r,\,x,\,r_{0})\,
\Ra\,=\,\nonumber \\
&\,&\,=\,\frac{Q_{l}(r)}{r}\,\psi_{\alpha\,l\,m}(t,\, r\,)\,,
\eeqar
with $Q_{l}$ as a $l$ multipole moment of the drop. 
The Schr\"{o}dinger equation \eq{2Ch3} now acquires the following form:
\beq\label{2Ch8}
\imath\,\frac{\partial }{\partial t}\,\psi_{\alpha\,l\,m}(t,\,r)\,= \,
\Le\,-\,\frac{1}{2 m}\,\Le\, \Delta_{r}\,-\,\frac{l\,(l\,+\,1)}{r^{2}}\,\Ra\,-\,\mu\,+\,
\frac{q^{2}}{r}\,Q_{l}(r)\,\Ra\,\psi_{\alpha\,l\,m}(t,\,r)\,.
\eeq
Representing the wave function as
\beq\label{2Ch9}
\psi_{\alpha\,l\,m}(t,\,r)\,=\,u_{\alpha\,l\,m}(r)\,e^{-\imath\,t\,\Le\,E\,-\,\mu\,\Ra}
\eeq
we obtain the Schr\"{o}dinger equation for the particle in the following form:
\beq\label{2Ch10}
\Le\, \frac{1}{2 m}\,\Delta_{r}\,-\,\frac{1}{2m}\,\frac{l\,(l\,+\,1)}{r^{2}}\,+\,E\,-\,
\frac{q^{2}}{r}\,Q_{l}(r)\,\Ra\,u_{\alpha\,l\,m}(r)\,=\,0\,.
\eeq
In general, we can not solve this equation without knowledge of the form of $f(r,\,x,\,r_{0})$ particles distribution function in \eq{2Ch6} integrals .
Nevertheless, we can guess the form of the function in the $r\,\leq\,r_{0}$ region of the drop, mostly interesting for us.
Indeed, at $r\,\gg\,r_{0}$, which is outside of the drop region, the potential \eq{2Ch7} is the usual Coulomb potential, but in the $r\,\leq\,r_{0}$ region, the 
situation is different. The existence of the drop requires the presence of some potential well at $r\,\leq\,r_{0}$ which will keep particles inside the drop
during some (very short) time and, therefore, it must be the potential's 
minimum present somewhere between $r\,=\,0$ and $r\,\propto\,r_{0}$. Hence, this minimum  is the indication of the
creation of the dense drop of finite size in the interaction system of interest and, consequently, we can write the potential energy from \eq{2Ch10} in this region as
\beq\label{2Ch11}
\frac{1}{2m}\,\frac{l\,(l\,+\,1)}{r^{2}}\,+\,\frac{Q_{l}(r)\,q^{2}}{r}\,\approx\,\frac{1}{2m}\,\frac{l\,(l\,+\,1)}{r^{2}_{min}}\,+\,
\frac{Q_{l}(r_{min})\,q^{2}}{r_{min}}\,+\,\frac{A_{l}(r_{min})\,q^{2}}{2\,r^{3}_{0}}\,(r\,-\,r_{min})^{2}\,,
\eeq
where we assumed that the potential energy acquires it's minimum at $r_{min}$, the $A_{l}(r_{min})$ here are the positive coefficients of the potential's expansion around this minimum.
This situation, in fact, is similar to the situation in the system of two-atoms molecules, see \cite{Mol} 
and thereafter, where two atoms are kept  inside some mutual potential well. 
Inserting \eq{2Ch11} expansion in \eq{2Ch10}, we obtain the following equation:
\beq\label{2Ch12}
\Le\, \frac{1}{2 m}\,\Delta_{r}\,+\,E\,-\,\frac{1}{2m}\,\frac{l\,(l\,+\,1)}{r^{2}_{min}}\,-\,Q_{l}(r_{min})\frac{q^{2}}{r_{min}}\,-\,A_{l}(r_{min})\,
\frac{q^{2}}{2\,r^{3}_{0}}\,(r\,-\,r_{min})^{2}\,\Ra\,u_{\alpha\,l\,m}(r)\,=\,0\,.
\eeq
The solution of this equation is similar to the solution of the Schr\"{o}dinger equation for the harmonic oscillator 
with energy levels defined by
\beq\label{2Ch121}
E^{'}\,=\,E\,-\,\frac{1}{2m}\,\frac{l\,(l\,+\,1)}{r^{2}_{min}}\,-\,Q_{l}(r_{min})\frac{q^{2}}{r_{min}}\,,
\eeq
and consequently the energy levels of the system are given by
\beq\label{2Ch13}
E_{n\,l}\,=\,\frac{1}{2m}\,\frac{l\,(l\,+\,1)}{r^{2}_{min}}\,+\,Q_{l}(r_{min})\frac{q^{2}}{r_{min}}\,+\,q\,\Le\,\frac{A_{l}(r_{min})}{m\,r_{0}^{3}}\,\Ra^{1/2}\,\Le\,n\,+\,1/2\,\Ra\,\,\,\,\,\,\,n\,=\,0,\,1,\,2,\,\ldots\,,
\eeq
with the wave functions
\beq\label{2Ch131}
u_{\alpha\,l\,m}(r)\,=c_{\alpha\,l\,m}\,\frac{e^{-(r\,-\,r_{min})^{2}/(2\,R^{2})}}{r\,\sqrt{R}}\,H_{n}((r\,-\,r_{min})\,/\,R)\,,
\eeq
where
\beq\label{2Ch14}
R\,=\,\Le\,\frac{r_{0}^{3}}{m\,A_{l}(r_{min})\,q^{2}}\,\Ra^{1/4}\,.
\eeq
We note, that the obtained solution is indeed similar to the solution of Schr\"{o}dinger equation for the two-atoms molecules, see \cite{Mol} for example.

\section{Conclusion}

\,\,\,\, In this note, we demonstrate that the spectrum of the charged non-relativistic particle in the dense charged drop has a quantum structure, see \eq{2Ch13}
and it is determined by three terms in the first mean field approximation.  The first term in \eq{2Ch13} can be considered as the quantized rotation energy of the drop, the second one
is the quantized electrostatic energy due to the miltipole moments of the charged drop. The third term in \eq{2Ch13} expression is the usual quantum corrections to the energy due to the 
oscillation of the particle inside the drop with 
eigenfrequencies determined by the form of the distribution function of the particles in the drop. The presence of this minimum is a necessary condition of the drop's creation, see
also \cite{Our}.
The values of these corrections have an additional  
degeneracy of energy levels defined by $l$ quantum number in comparison to the ordinary quantum oscillator.
In this formulation, the considered problem is similar to the problem of the description of system of two-atom molecule, see \cite{Mol}
\footnote{We note, that the proposed approach can be used also for the description of bound states created at low energy interactions as well, 
we plan to investigate this subject in a separate publication.}. The further development of the approach can include 
the consideration of higher orders of mean mean field approximation for the system and introducing the kinetic equation for the \eq{1Ch3} distribution function coupled to the Hamiltonian, 
these problems we plan to consider in the 
following publications.
 
  We conclude, that our model can be useful for the clarification of the spectrum of the produced particles which is influenced by the
quantum-mechanical properties of the QCD fireball.	
We believe, that this approach  
will provide the connection
between the data, obtained in high-energy collisions of protons and nuclei in the LHC and RHIC experiments \cite{Fluid,Fluid1,Fluid2,BNL,BNL3}, and microscopic fields inside the collision region.


\end{document}